\documentclass[%
reprint,
 amsmath,amssymb,
 aip,
floatfix,
]{revtex4-2}

\usepackage{comment}

\usepackage[utf8]{inputenc}
\usepackage[T1]{fontenc}
\usepackage{mathptmx}
\usepackage{etoolbox}

\usepackage{subfig}
\usepackage{wasysym}
\usepackage{stmaryrd}
\usepackage{amsfonts}
\usepackage{xcolor}

\usepackage{graphicx}
\usepackage{dcolumn}
\usepackage{bm}
\usepackage{hyperref}

\allowdisplaybreaks

\makeatletter
\def\@email#1#2{%
 \endgroup
 \patchcmd{\titleblock@produce}
  {\frontmatter@RRAPformat}
  {\frontmatter@RRAPformat{\produce@RRAP{*#1\href{mailto:#2}{#2}}}\frontmatter@RRAPformat}
  {}{}
}%
\makeatother

\begin{document}


\title[Proposition of extension of models relating rheology and microscopic structure with a double fractal structure]{Proposition of extension of models relating rheological quantities and microscopic structure through the use of a double fractal structure}

\author{Louis-Vincent Bouthier}
\email{louis.bouthier@mines-paristech.fr}
\author{Romain Castellani}%
\author{Elie Hachem}
\author{Rudy Valette}
\affiliation{%
 Groupe CFL, CEMEF, Mines Paris, PSL Research University, 1 Rue Claude Daunesse, 06904 Sophia Antipolis, France
}%

%




\date{\today}

\begin{abstract}

Colloidal suspensions and the relation between their rheology and their microstructure is investigated. The literature showed great evidence of the relation between rheological quantities and particle volume fraction, ignoring the influence of the cluster. We propose to extend previous models using a new double fractal structure which allows, first, to recover the well-known models on the case of percolated system and, second, to capture the influence of the cluster size. This new model emphasises the necessity of such structure to account for recent experimental results. Then, the model is compared with data coming from the literature and shows close agreement.

\end{abstract}

\maketitle





Colloidal suspensions are composed by a liquid medium containing particles, which aggregate in clusters through attractive forces. Such suspensions have been of great interest for recent papers \cite{Gibaud2020a,Dages2021,
Kimbonguila2014,
Wessel1992,
Nguyen2011}. The particles can be made of alumina \cite{Nguyen2011,Waite2001,Schilde2011,Sauter2008,Mahbubul2014}
, polystyrene \cite{Okubo1995}, silica \cite{Mondragon2012},  titania \cite{Fazio2008} or with other materials. Besides, the solvent can be water \cite{Nguyen2011,Vassileva2005}, a mineral oil \cite{Gibaud2020a,Dages2021,Keshavarz2021} or another liquid. The attractive forces
are also diverse with pure Van der Waals interaction \cite{Visser1972,Hartley1985}, Derjaguin, Landau, Verwey, and
Overbeek like interaction \cite{Keshavarz2021,Gibaud2020a,Jiang2020} or capillary bridges \cite{Vassileva2005,Rahman2019,Kralchevsky2001a}
. Several types of behaviour can be assessed like shear-thickening \cite{Lee2021} or other viscosity dependence \cite{Mwasame2016}. Even with the broad literature on this topic, the precise influence of the clusters on the rheological properties of the suspension remains an open question. The clusters are mainly showing a fractal behaviour with ramified structures and varying fractal dimension $D$ or chemical dimension $d$\cite{Herrmann1984,Grassberger1992a}
.

Several studies tried to understand the micromechanisms creating the clusters \cite{
Kimbonguila2014,Eggersdorfer2010,
Fielding2020,
Wessel1992,Kantor1984a}, 
relating their influence on the rheology to well-known models \cite{Macosko1994,Herschel1926,
Bingham1922}. 
Nevertheless, the complete coupling between the microscopic scale and the macroscopic scale and the true nature of the fractal clusters remains unknown.

A wide part of the literature often identifies a relation for the storage modulus $G'$ and the linearity limit of strain $\gamma_{\mathrm{NL}}$ with the particle volume fraction $\phi$ of the following form

\begin{align}
    G'&\propto\phi^\mu\\
    \gamma_\mathrm{NL}&\propto\phi^\nu
\end{align}
with $\mu\in\left[3,5\right]$ and $\nu\in\left[-2,1\right]$ \cite{
Shih1990, Mellema2002,
Marangoni2000,
Wu2001}. The papers are often linking these exponents to intrinsic structure parameters like the fractal dimension $D$, the chemical dimension $d$, the proportion of bending or stretching $\epsilon$, the proportion of strong-link regime or weak-link regime $\alpha$ and the dimension of the Euclidean space. Nevertheless, these theoretical models do not take into account the possibility for the system, composed mainly by clusters (or flocs, or aggregates, depending on the terminology), to exhibit variation of structure, particularly in terms of size of clusters. One of the common hypothesis of all the models is that the size of the clusters $\ell$ is uniquely determined by the particle size $a$, the dimension of the Euclidean space $\dim$, $D$ and $\phi$ through $\phi_\mathrm{eff}=\phi\left(\ell/a\right)^{\dim-D}=1$ \footnote{Measuring a cluster size is not trivial due to technical/experimental difficulties to reach it and to a possible spread of size coming from a statistical distribution of size. Nevertheless, experimental tools and techniques exist to reach this kind of information \cite{Weitz1984,Weitz1985,Wagner1990} and statistical description may cover various needs \cite{Banasiak2020a,
Sorensen1987}
}. However, recent papers \cite{Gibaud2020a,Dages2021}
have shown the importance of the microstructure of clusters on the storage modulus $G'$ in particular with a constant particle volume fraction $\phi$. The storage modulus must then depend on the size of the clusters, or more generally on the microstructure.


The objective of this paper is an attempt to describe a new model of suspension to get rheological quantities as well as recovering well-known models. This paper tries to propose extensions of the previous models, assuming a double fractal structure and to invite other research to investigate, maybe, this opportunity 
. Therefore, 
first, the description of the model is presented. Afterwards 
 the results are showed, discussed according to the literature, and some conclusions are drawn.

\label{sec:Theory}

\begin{figure}
    \centering
    \includegraphics[width=\columnwidth]{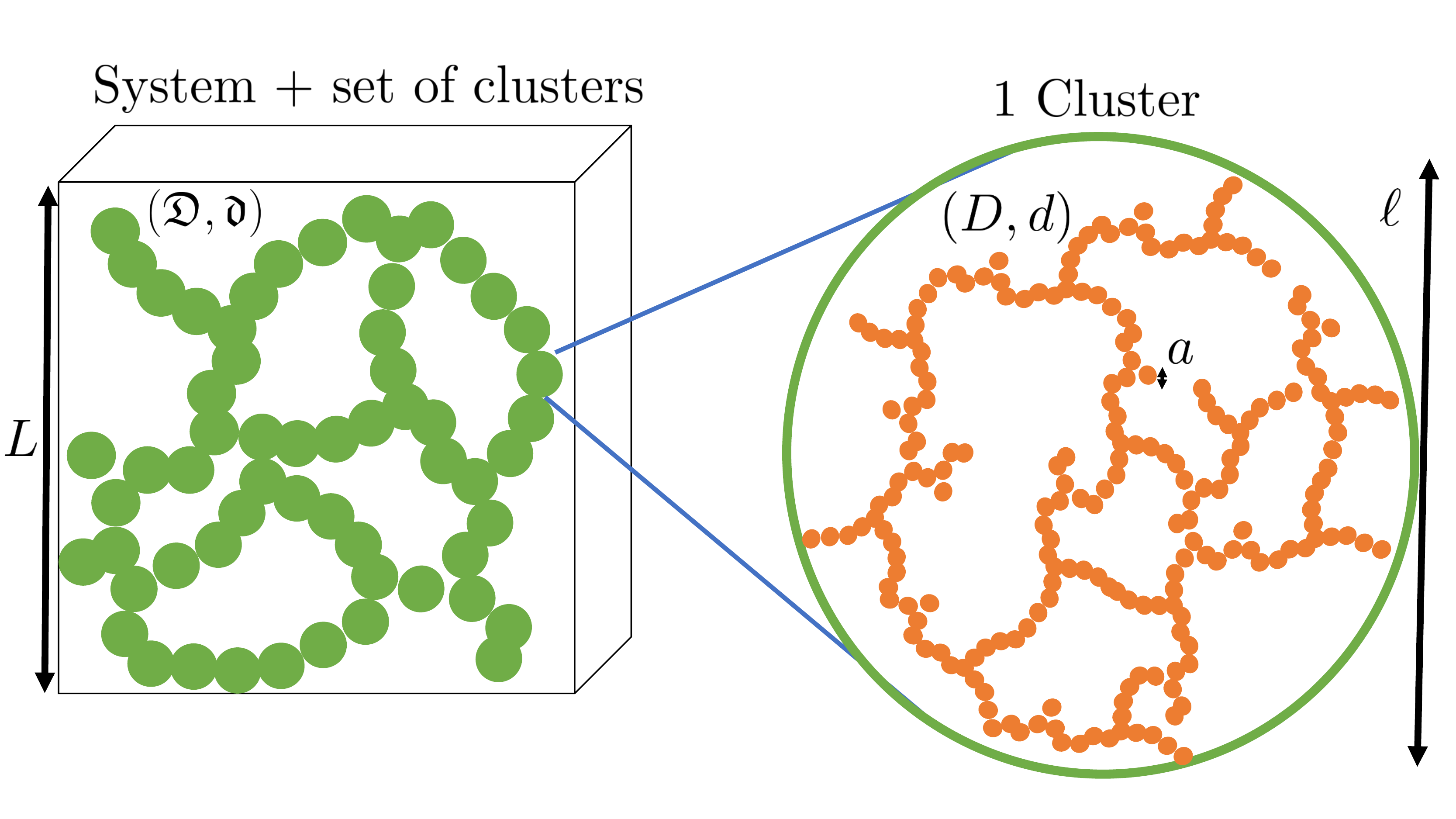}
    \caption{Sketch of the double fractal structure transmitting stresses in the system, in the clusters and between the particles}
    \label{fig:StorageModulus}
\end{figure}

To build the storage modulus of a colloidal suspension, let us consider a percolated system of macroscopic size $L$ (see Fig. \ref{fig:StorageModulus}) in a Euclidean space of dimension $\dim$. The gel is composed of particles of size $a$ with a volume fraction $\phi$ aggregated in clusters of size $\ell$ with an interaction potential $U$ and a distance of interaction $\delta$ (see Refs.~ \onlinecite{Marshall2014,Kimbonguila2014,Eggersdorfer2010}
). The size of the cluster $\ell$ can be determined by multiple factors (volume fraction, interaction potential, external solicitation, ...). Some models demonstrate this kind of relationship \cite{Sorensen1987,Ruan2020,Kimbonguila2014}
but one will consider this size here as a  variable. Then, compared to previous models \cite{
Shih1990,Mellema2002,
Marangoni2000,
Wu2001}, this size is not assumed to follow $\phi\left(\ell/a\right)^{\dim-D}=1$ ($D$ being the fractal dimension) 
. Indeed, in this paper, one assumes the existence of a double fractal structure in the system, each one having two parameters, as depicted on Fig. \ref{fig:StorageModulus}. This assumption, discussed in the rest of the paper, is crucial and arises from both the existence of size of clusters $\ell$ and of the percolated elasticity: the coexistence of both latter ones implies the former assumption. The first structure relates the particles and the clusters, having  a fractal dimension $D$ and a chemical dimension or shortest path dimension $d$. $d$ is also the dimension of the elastic backbone and is between 1.1 and 1.4 \cite{Herrmann1984,Grassberger1992a}
. The second structure is between the clusters and the macroscopic system, having a fractal dimension $\mathfrak{D}$ and a chemical dimension $\mathfrak{d}$. For now, there are no more assumptions about the value of $\left(\mathfrak{D},\mathfrak{d}\right)$ respectively to $\left(D,d\right)$. The assumption of double structure is related to the fact that, in a percolated system, the clusters are gathered in a somewhat structure which covers the whole system: to be able to handle elastic deformation, the volume 
is spanned with a network of clusters which are themselves composed of  particles that span space. Indeed, this assumption is typically supported by  small angle scattering measurements \cite{Gibaud2020a,Dages2021,
Weitz1984,Weitz1985}. If this double structure would not exist, the identification of a cluster size would not show any difference between the sup-cluster and the sub-cluster structure, then would not allow to identify properly a size $\ell$. In other words, the double fracture structure is required to identify a cluster size. If this structure was regular such as a generalized cubic network, then we would have $\left(\mathfrak{D},\mathfrak{d}\right)=\left(\dim,1\right)$. However one leaves the possibility to have a more complex structure above the clusters. This will impact the capacity of the gel to change its storage modulus according to the microscopic structure. 

Our expression of the storage modulus $G'$ follows the reasoning proposed in Refs.~\onlinecite{
Shih1990,Mellema2002,
Marangoni2000,
Wu2001} 
in which assumptions of springs in series in a fractal structure are made. Therefore, it is necessary to know the macroscopic stiffness $K$ of the system, which reads, as a first approach $G'=K/L^{\dim-2+2\epsilon\alpha}$, where 
$\epsilon\in\left[0,1\right]$ (considering the range between pure stretching $\epsilon=0$ and pure bending $\epsilon=1$) and $\alpha\in\left[0,1\right]$ (considering the range between strong-link regime $\alpha=1$ and weak-link regime $\alpha=0$\footnote{In the case of weak-link regime, the values of $\epsilon$, $d$ and $\mathfrak{d}$ are useless due to irrelevance of information of internal elastic backbone.} \cite{
Shih1990,Mellema2002,
Marangoni2000,
Wu2001}). The power $\dim-2+2\epsilon\alpha$ for the macroscopic size $L$ is an attempt to take into account the different regimes proposed in the previous models in the literature and modify the dimensions of $K$ from a linear spring to a torsion spring. Then $K$ is linked to the stiffness of each cluster $k_c$ as if the clusters were in series, which brings $K=k_c\left(\ell/L\right)^\mathfrak{d\alpha}$. The distinction between strong-link and weak-link regime is blatant because, in the former case, all clusters of size $\ell$ are contributing to the global stiffness, whereas, for the latter, only the extreme clusters are playing a role, thus only one cluster appears. Furthermore, the stiffness $k_c$ is related to the stiffness between the particles $k_p$ having a similar behaviour of springs in series, which leads to $k_c=k_pa^{2\epsilon\alpha}\left(a/\ell\right)^{d\alpha}$. The power $d\alpha$ is directly related to the difference between strong-link and weak-link regimes, where either all the particles contribute in the series or only the particles at the boundary. Also, the factor $a^{2\epsilon\alpha}$ relates the particle interaction stiffness to more general movement from stretching to bending through curved paths. Finally, the stiffness of each particle bond $k_p$ is linked to the interaction potential with $k_p=U\delta^{-2}$. Moreover, because of the fractal structure of the system, one has $\phi_\mathrm{eff}=\phi\left(\ell/a\right)^{\dim-D}=\left(\ell/L\right)^{\dim-\mathfrak{D}}$. This finally brings

\begin{equation}
    G'=\frac{U}{a\delta^2}\phi^{\frac{\dim-1+f\left(\mathfrak{d}\right)}{\dim-\mathfrak{D}}}\left(\frac{\ell}{a}\right)^{\frac{\dim-D}{\dim-\mathfrak{D}}\left(\dim-1+f\left(\mathfrak{d}\right)\right)
    -\dim+1-f\left(d\right)}\label{eq:StorageModulus}
\end{equation}
with $f\left(x\right)=\alpha\left(2\epsilon+x\right)-1$.

A first comment for this equation is that assuming a percolated system with $\phi_\mathrm{eff}\approx1$, one recovers the known behaviour $G'\propto\phi^{\frac{\dim-1+f\left(d\right)}{\dim-D}}$ \cite{
Mellema2002,
Marangoni2000,
Shih1990,
Wu2001}.  Another comment is that if $\mathfrak{D}=\dim$ there is no opportunity for the system to let the clusters influence the rheology: thus this assumption of $\mathfrak{D}\ne\dim$ is absolutely necessary. This becomes relevant when one considers a generalized cubic network system where there is no particular reason for the network to break under a change of structure. Also, if $\mathfrak{D}$ would be equal to $\dim$,  $L$ would not have any influence on the rheology which is definitely not true, considering that the assembly of clusters is bringing the overall stiffness of the system: a higher or lower number of clusters may bring differences in the macroscopic stiffness. About the function $f$, depending on the values of $\alpha\in\left[0,1\right]$, $\epsilon\in\left[0,1\right]$, $d\in\left[1.1,1.4\right]$ and $\mathfrak{d}\in\left[1.1,1.4\right]$, it is varying between -1 and 2.4 approximately 
.

This approach can also be extended to express the linearity limit of strain $\gamma_\mathrm{NL}$. For instance, using the same approach shown in Refs.~\onlinecite{Shih1990,
Mellema2002,
Marangoni2000,
Wu2001}, one assumes that $\sigma_y=G'\gamma_\mathrm{NL}$ should not depend on $d$, $\mathfrak{d}$, $\epsilon$ and $\alpha$ because it is intrinsic and depends mainly on the pair interaction potential. Also, following previous approaches with their set of hypothesis, $\sigma_y\propto\ell^{1-\dim}$ and $\sigma_y\propto\phi^{\frac{\dim-1}{\dim-D}}$. Finally, this reads
\begin{equation}
    \gamma_\mathrm{NL}=\frac{\delta}{a}\phi^{\frac{-f\left(\mathfrak{d}\right)}{\dim-\mathfrak{D}}}\left(\frac{\ell}{a}\right)^{-\frac{\dim-D}{\dim-\mathfrak{D}}f\left(\mathfrak{d}\right)\\
    +f\left(d\right)}. \label{eq:YieldStrain}
\end{equation}

Assuming again a percolated system with $\phi_\mathrm{eff}\approx1$, one recovers $\gamma_\mathrm{NL}\propto\phi^{\frac{-f\left(d\right)}{\dim-D}}$ \cite{
Mellema2002,
Marangoni2000,
Shih1990,
Wu2001}. 
Hence, the yield stress $\sigma_y$ may be obtain thanks to Eq. (\ref{eq:StorageModulus}) and Eq. (\ref{eq:YieldStrain}) to get 
\begin{equation}
    \sigma_y=\frac{U}{a^{2}\delta}\phi^{\frac{\dim-1}{\dim-\mathfrak{D}}}\left(\frac{\ell}{a}\right)^{\frac{\left(\dim-1\right)\left(\mathfrak{D}-D\right)}{\dim-\mathfrak{D}}} \label{eq:yieldstress}
\end{equation}
which brings the well-known behaviour $\sigma_y\propto\phi^{\frac{\dim-1}{\dim-D}}$ in a percolated system with $\phi_\mathrm{eff}\approx1$ \cite{
Mellema2002,
Marangoni2000,
Shih1990,
Wu2001}. It is clear from Eq. (\ref{eq:yieldstress}) that if $\mathfrak{D}=D$, there is no more influence of the size of clusters $\ell$ on the yield stress. More generally, from Eqs. (\ref{eq:StorageModulus}), (\ref{eq:YieldStrain}) and (\ref{eq:yieldstress}), if $\left(\mathfrak{D},\mathfrak{d}\right)=\left(D,d\right)$ as a classical percolated system, the size of the clusters $\ell$ will not have any influence on the rheology. 

This consequence can be interpreted as a critical phenomenon because it is necessary to have a difference between $\left(\mathfrak{D},\mathfrak{d}\right)$ and $\left(D,d\right)$. Heterogeneity in the double fractal structure leads to fragility and to the possibility of breakage. Also, depending on the values of $\left(D,d,\alpha,\epsilon,\mathfrak{D},\mathfrak{d}\right)$, it is clear that the evolution of $G'$ and $\sigma_y$ according to $\phi$ is always increasing, the evolution of $\gamma_\mathrm{NL}$ according to $\phi$ may be either increasing or decreasing, and the evolution of $G'$, $\gamma_\mathrm{NL}$ and $\sigma_y$ according to $\ell/a$ may be either increasing or decreasing. Thus, this model allows a variety of systems with different phenomenology, depending on the real behaviour of the material.

\begin{figure}
    \centering
    \includegraphics[width=\columnwidth]{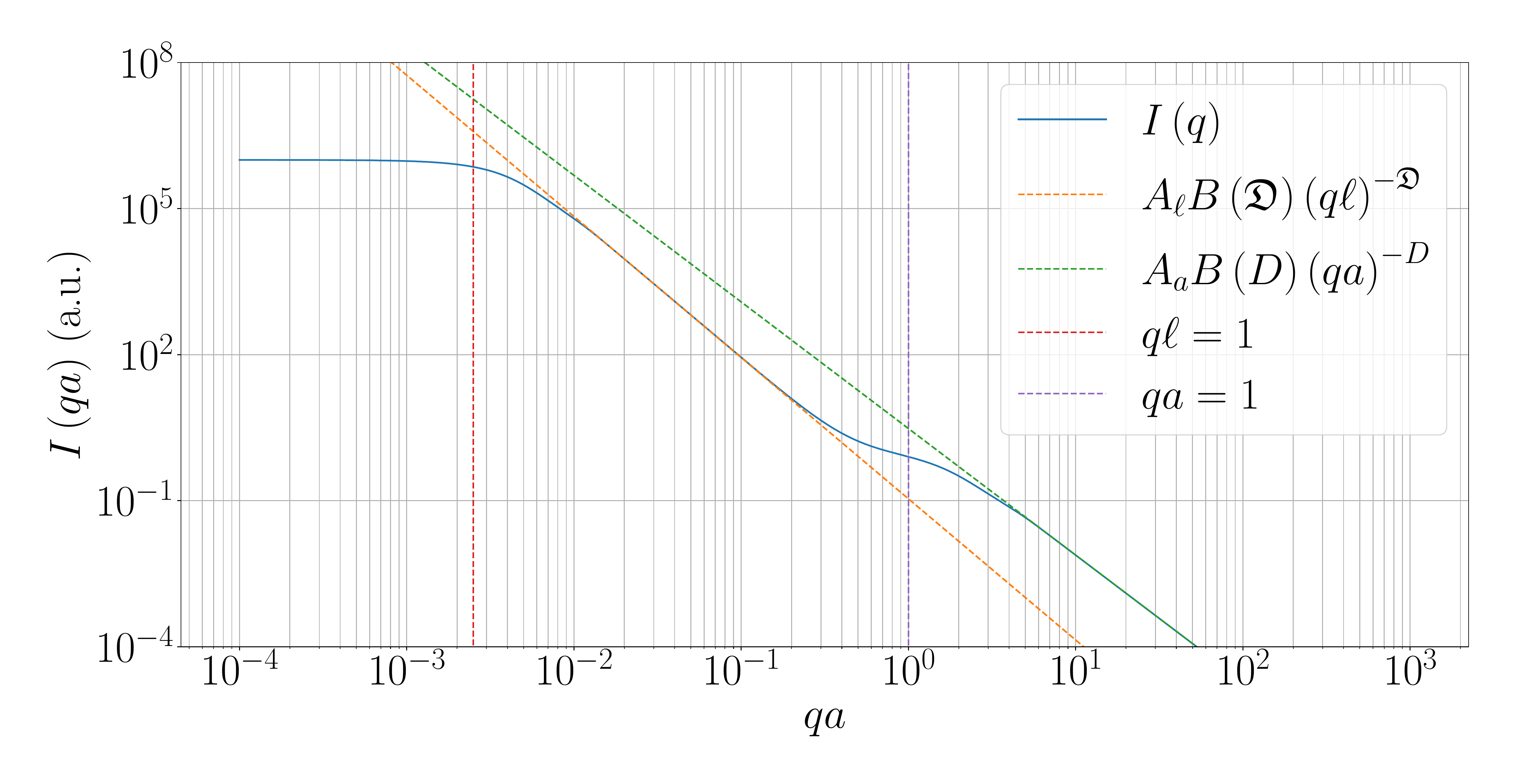}
    \caption{Example of intensity spectrum extracted from Eq. (\ref{eq:Intensity}) with $A_\ell=10^6\mathrm{a.u.}$, $A_a=1\mathrm{a.u.}$, $\ell/a=400$, $D=2.6$ and $\mathfrak{D}=2.9$}
    \label{fig:Intensity}
\end{figure}

One can also try to identify the intensity spectra of such a system which can be directly inferred through \cite{Gibaud2020a,
Dages2021,Sorensen2001,Hammouda2010} 
\begin{multline}
    \label{eq:Intensity}
    I\left(q\right)=A_\ell\exp\left(-\frac{q^2\ell^2}{3}\right)\\
    +\frac{A_\ell B\left(\mathfrak{D}\right)}{\left(q\ell\right)^{\mathfrak{D}}} \exp\left(-\frac{q^2a^2}{3}\right)\mathrm{erf}\left(\frac{q\ell}{\sqrt{6}}\right)^{3\mathfrak{D}}\\
    +A_a\left[\exp\left(-\frac{q^2a^2}{3}\right)+\frac{B\left(D\right)}{\left(qa\right)^{D}}\mathrm{erf}\left(\frac{qa}{\sqrt{6}}\right)^{3D}\right]
\end{multline}
with 
\begin{equation}
    B\left(\Delta\right)=\Delta\left(\frac{3\Delta^2}{\left(2+\Delta\right)\left(1+\Delta\right)}\right)^{\frac{\Delta}{2}}\Gamma\left(\frac{\Delta}{2}\right),
\end{equation}
$A_a$ and $A_\ell$ being empirical constants to fit, $q$ the wave vector norm, $\mathrm{erf}$ the error function and $\Gamma$ the Gamma function. The Guinier regime at low values of $q$\cite{Sorensen2001} is recovered with the exponential terms in Eq. (\ref{eq:Intensity})  and the other terms with $\mathrm{erf}\left(x/\sqrt{6}\right)^3/x\underset{x\to0}{\sim}x^2\left(2/3\pi\right)^{3/2}$. The fractal slope is recovered with the fast decaying exponential, the limit value of the error function towards infinity and the power law decrease involving the fractal dimension. One thing to note to properly identify the intensity spectra is that the range investigated $\left[q_{\min},q_{\max}\right]$ and the system should verify $aq_{\max}>2\pi$, $\ell \apprge 50a$, $50q_{\min}\ell\apprle2\pi$ which leads to $q_{\max}/q_{\min}>2,500$ which is rather large in terms of orders of magnitude. Few methods, particularly scattering techniques, and physical systems allow to reach such a broad range for the wave vector. One example of spectrum is given on Fig. \ref{fig:Intensity} where it is easy to identify $a$ with the rupture of slope on the right, $\ell$ with the rupture of slope on the left, $D$ with the power law slope on the right and $\mathfrak{D}$ with the power law slope on the left.

A development of the model for multiple levels may be proposed\footnote{The previous model can be extended to $n\in\mathbb{N}\setminus\left\{0,1\right\}$ number of steps depending on the type of the considered system. The more  steps, the larger quantity of parameter to identify. Therefore, the construction may be interesting but should remain seldom due to broad variability of parameters. As a snapshot, giving a set $\left(\ell_i,D_i,d_i\right)_{i\in\left\llbracket1,n\right\rrbracket}$ of sizes, fractal dimension and chemical dimension, $n\in\mathbb{N}\setminus\left\{0,1\right\}$, $\ell_0=a$, $\ell_n=L$, Eqs. (\ref{eq:StorageModulus}), (\ref{eq:YieldStrain}) and (\ref{eq:yieldstress}) read respectively
\begin{align}
    G'&=\frac{U}{a\delta^{2}}\phi^{\frac{\dim-1}{\dim}}\prod_{i=1}^{n}\left(\frac{\ell_{i-1}}{\ell_{i}}\right)^{f\left(d_{i}\right)+D_{i}\frac{\dim-1}{\dim}}\\
\gamma_{\mathrm{NL}}&=\frac{\delta}{a}\prod_{i=1}^{n}\left(\frac{\ell_{i-1}}{\ell_{i}}\right)^{-f\left(d_{i}\right)}\\
\sigma_{y}&=\frac{U}{a^{2}\delta}\phi^{\frac{\dim-1}{\dim}}\prod_{i=1}^{n}\left(\frac{\ell_{i-1}}{\ell_{i}}\right)^{D_{i}\frac{\dim-1}{\dim}}\\
\phi&=\left(\frac{a}{L}\right)^{\dim}\prod_{i=1}^{n}\left(\frac{\ell_{i-1}}{\ell_{i}}\right)^{-D_{i}}.
\end{align}
In order to recover an expression similar to the previous ones in terms of the volume fraction, one needs to integrate \emph{partial} volume fraction taking into account the volume fraction into a cluster of a certain size. Considering these previous expressions, one can extend to a continuum of length scales with two functions $\ell\mapsto d\left(\ell\right)$ and $\ell\mapsto D\left(\ell\right)$ for the chemical dimension and the fractal dimension respectively to get
\begin{align}
    G'&=\frac{U}{a\delta^{2}}\phi^{\frac{\dim-1}{\dim}}\exp\left(-\int_{a}^{L}\frac{1}{\ell}\left(f\left(d\left(\ell\right)\right)+\frac{\dim-1}{\dim}D\left(\ell\right)\right)\,\mathrm{d}\ell\right)\\
\gamma_{\mathrm{NL}}&=\frac{\delta}{a}\exp\left(\int_{a}^{L}\frac{f\left(d\left(\ell\right)\right)}{\ell}\,\mathrm{d}\ell\right)\\
\sigma_{y}&=\frac{U}{a^{2}\delta}\phi^{\frac{\dim-1}{\dim}}\exp\left(-\frac{\dim-1}{\dim}\int_{a}^{L}\frac{D\left(\ell\right)}{\ell}\,\mathrm{d}\ell\right)\\
\phi&=\exp\left(\int_{a}^{L}\frac{D\left(\ell\right)}{\ell}\,\mathrm{d}\ell\right)\left(\frac{a}{L}\right)^{\dim}.
\end{align}

With these expressions, it is easy to recover the model with one fractal structure or two fractal structure taking constant functions or two-step constant function respectively over $\left[a,L\right]$.}.

\label{sec:Results}


One can look at the assessment of the storage modulus, the linear limit of strain and the yield stress according to literature results. Using Eqs. (\ref{eq:StorageModulus}), (\ref{eq:YieldStrain}) and (\ref{eq:yieldstress}) and Tab. \ref{tab:LiteratureResultsStorage}, the values of $d$ and $\mathfrak{d}$ have been assumed close to the lower boundaries of their range (i.e. $\left[1.1,1.4\right]$), $\mathfrak{D}$ has been chosen close to $D$ remaining higher to have an increase of the storage modulus $G'$ according to $\ell/a$, and $\phi$ is chosen according to the structure of the fractal particles of the carbon black. The theoretical value of storage modulus is $G'_\mathrm{th}=0.96\mathrm{kPa}$ which is close to the experimental value $G'_{\exp}=1.2\mathrm{kPa}$. Also, the theoretical value of the limit of linearity strain $\gamma_\mathrm{NL}^\mathrm{th}=0.34\%$ is also close to the experimental value $\gamma_\mathrm{NL}^{\exp}=1\%$. Finally, the theoretical value of the yield stress $\sigma_y^\mathrm{th}=3.4\mathrm{Pa}$ is close to the experimental value $\sigma_y^{\exp}=12\mathrm{Pa}$ too. 

\begin{figure}
    \centering
    \includegraphics[width=\columnwidth]{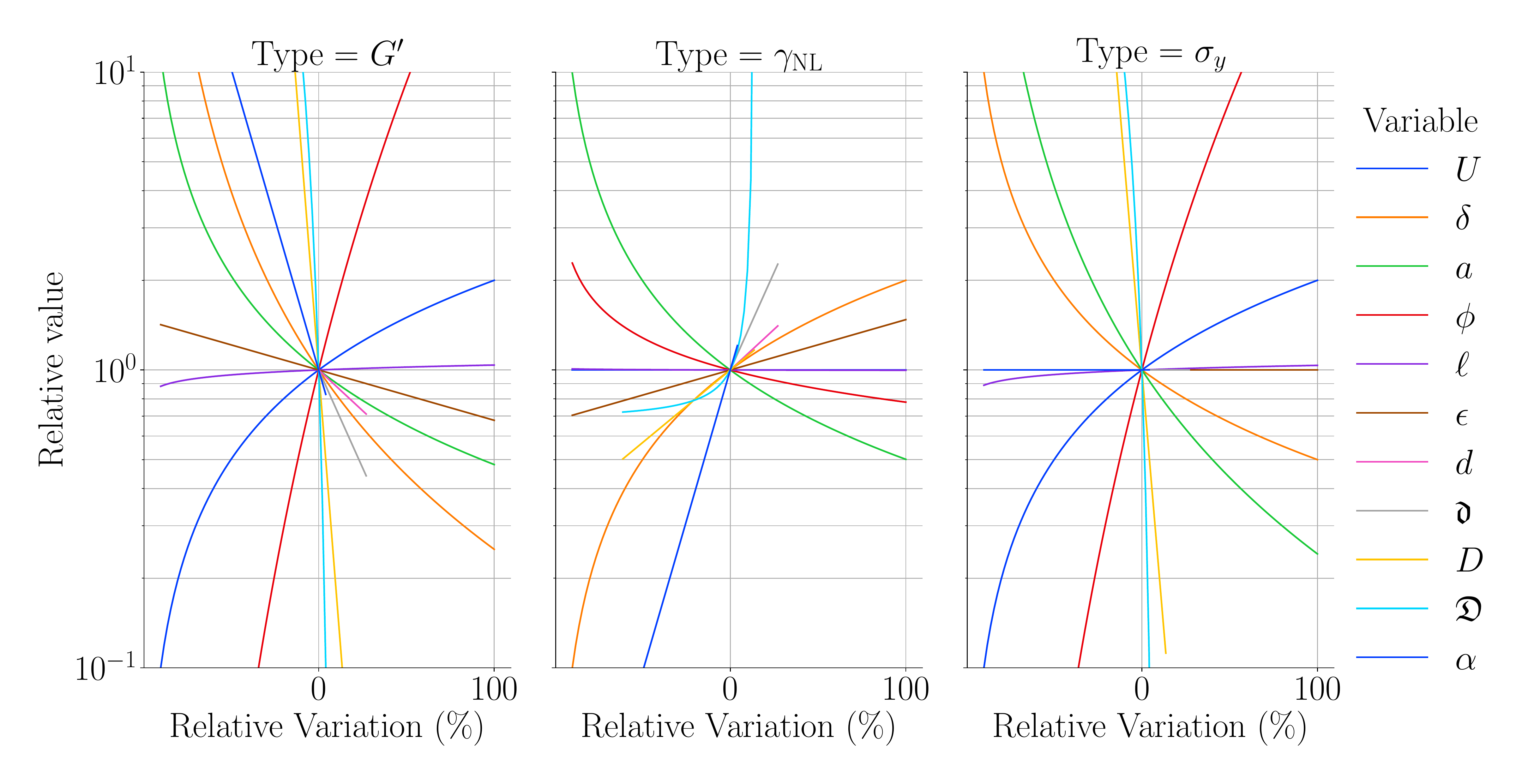}
    \caption{Sensitivity study of $G'$, $\gamma_\mathrm{NL}$ and $\sigma_y$ from left to right respectively, according to Eqs. (\ref{eq:StorageModulus}), (\ref{eq:YieldStrain}) and (\ref{eq:yieldstress}) following the relative variation of the parameters around the value given in Tab. \ref{tab:LiteratureResultsStorage}. The conditions $\left(d,\mathfrak{d}\right)\in\left[1.1,1.4\right]^2$, $\left(D,\mathfrak{D}\right)\in\left]2,3\right[^2$, $\epsilon\in\left[0,1\right]$, $\alpha\in\left[0,1\right]$ and $\dim=3$ have been implemented.}
    \label{fig:SensitivityG}
\end{figure}

Convincingly, a sensitivity study can be carried leading to Fig. \ref{fig:SensitivityG}. The most critical parameters in this study are then $\phi$, $\mathfrak{D}$, $\epsilon$, $D$ and $\delta$ as suggested by Eqs. (\ref{eq:StorageModulus}), (\ref{eq:YieldStrain}) and (\ref{eq:yieldstress}). Experimental measures of these parameters are then of particular interest and needs to bring accurate values.

\begin{table}
    \centering
    \begin{ruledtabular}
    \begin{tabular}{ld}
        Source & Refs.~\cite{Gibaud2020a,Dages2021} \\
        \hline
        $G'_\mathrm{exp}$ (kPa) & 1.2\\
        $\gamma_\mathrm{NL}^{\exp}$ (\%) & 1\\
        $\sigma_y^{\exp}$ (Pa) & 12\\
        \hline
        $G'_\mathrm{th}$ (kPa) & 0.96\\
        $\gamma_\mathrm{NL}^\mathrm{th}$ (\%) & 0.34\\
        $\sigma_y^\mathrm{th}$ (Pa) & 3.4\\
        \hline
        $\phi$ (\%)& 20 \\
        $a$ (nm)& 150 \\
        $U$ ($kT$) & 20 \\
        $\ell$ (nm) & 500\\
        $\delta$ ($\mathring{\mathrm{A}}$) & 3 \\
        $D$ & 2.6 \\
        $d$ & 1.1 \\
        $\mathfrak{D}$ & 2.61 \\
        $\mathfrak{d}$ & 1.1 \\
        $\epsilon$ & 0.05 \\
        $\alpha$ & 0.95 \\
        $\dim$ & 3\\
    \end{tabular}
    \end{ruledtabular}

    \caption{Results and parameters to assess the rheological properties in Refs.~ \onlinecite{Gibaud2020a,Dages2021}. $G'_\mathrm{exp}$, $\gamma_\mathrm{NL}^{\exp}$ are the direct measure of the storage modulus at strain $\gamma\to0$ and the limit strain of linearity of the carbon black particle suspension, respectively. Also,  $\sigma_y^{\exp}=G'_\mathrm{exp}\gamma_\mathrm{NL}^{\exp}$. Then, $G'_\mathrm{th}$, $\gamma_\mathrm{NL}^\mathrm{th}$ and $\sigma_y^\mathrm{th}$ are the estimated storage modulus according to Eq. (\ref{eq:StorageModulus}), the limit strain of linearity according to Eq. (\ref{eq:YieldStrain}) and the yield stress according to Eq. (\ref{eq:yieldstress}), respectively, with the other parameters.}
    \label{tab:LiteratureResultsStorage}
\end{table}

After having presented several types of results coming from the literature and experiments, some main discussions and conclusions can be drawn. 
There is no doubt about the relation between the microscopic structure and the macroscopic behaviour. In this paper, one  tried to extend common knowledge about scaling laws between the rheological quantities and the particle volume fraction taking into account the influence of the size of the clusters. One hopes that this proposition may help to consider some discrepancies between experimental data, simulation data and the most usual models.


\appendix


\label{sec:Multiple}

\bibliography{Bibliographie_these}
\end{document}